\documentclass[10pt,superscriptaddress,nofootinbib,notitlepage,twocolumn,pra]{revtex4-2}

\usepackage{amsthm,amsmath,amssymb}
\usepackage{mathrsfs}
\usepackage{appendix}
\usepackage{amstext}
\usepackage{graphicx}
\usepackage{url}
\usepackage{float}
\usepackage{bm}
\usepackage{ifthen}
\usepackage[usenames,dvipsnames]{color}
\usepackage{mathrsfs}
\usepackage[colorlinks=true, citecolor=blue, urlcolor=black]{hyperref}
\usepackage{dcolumn}
\usepackage{natbib}
\usepackage{booktabs}
\usepackage{hyperref}
\usepackage{color}
\usepackage{multirow}
\usepackage{mathtools}
\usepackage{lipsum}

\newcommand{\bra}[1]{\mbox{$\left \langle #1 \right|$}}
\newcommand{\ket}[1]{\mbox{$\left| #1 \right\rangle$}}
\newcommand{\etal}{\mbox{$et$ $al$. }}
\newcommand{\aH}[1]{\mbox{$a^{H\dagger}_{#1}$}}
\newcommand{\aV}[1]{\mbox{$a^{V\dagger}_{#1}$}}

\begin{document}
	
	\title{Breaking Rate-Distance Limitation of Measurement-Device-Independent Quantum Secret Sharing}
	
    \author{Chen-Long Li}\thanks{These authors contributed equally to this work}
	\affiliation{National Laboratory of Solid State Microstructures and School of Physics, Collaborative Innovation Center of Advanced Microstrucstures, Nanjing University, Nanjing 210093, China}
	\author{Yao Fu}\thanks{These authors contributed equally to this work}	
	\affiliation{Beijing National Laboratory for Condensed Matter Physics and Institute of Physics, Chinese Academy of Sciences, Beijing 100190, China}
	\author{Wen-Bo Liu}
	\author{Yuan-Mei Xie}
	\author{Bing-Hong Li}
	\author{Min-Gang Zhou}
	\affiliation{National Laboratory of Solid State Microstructures and School of Physics, Collaborative Innovation Center of Advanced Microstrucstures, Nanjing University, Nanjing 210093, China}
	\author{Hua-Lei Yin}\email{hlyin@nju.edu.cn}
	\author{Zeng-Bing Chen}\email{zbchen@nju.edu.cn}
	\affiliation{National Laboratory of Solid State Microstructures and School of Physics, Collaborative Innovation Center of Advanced Microstrucstures, Nanjing University, Nanjing 210093, China}
\date{\today}
	
	\begin{abstract}
		Currently most progresses on quantum secret sharing suffer from rate-distance bound, and thus the key rates are limited.
        In addition to the limited key rate, the technical difficulty and the corresponding cost together prevent large-scale deployment. 
		Furthermore, the performance of most existing protocols is analyzed in the asymptotic regime without considering participant attacks. 
		Here we report a measurement-device-independent quantum secret sharing protocol with improved key rate and transmission distance.
	    Based on spatial multiplexing, our protocol shows it can break rate-distance bounds over network under at least ten communication parties.
		Compared with other protocols, our work improves the secret key rate by more than two orders of magnitude and has a longer transmission distance.
		We analyze the security of our protocol in the composable framework considering participant attacks and evaluate its performance in the finite-size regime.
		In addition, we investigate applying our protocol to digital signatures where the signature rate is improved more than $10^7$ times compared with existing protocols.
		We anticipate that our quantum secret sharing protocol will provide a solid future for multiparty applications on the quantum network.
	\end{abstract}
	
	\maketitle
	\section{Introduction}
	A network with quantum resources has benefits in both computing enabled by quantum computation~\cite{arute2019quantum,zhong2020quantum,zhong2021phase,wu2021strong,liu2021rigorous,zhou2022experimental} and secure communication enabled by quantum key distribution~\cite{gisin2002quantum, bennett1992communications}.
	Apart from quantum key distribution, in the realm of quantum communication quantum secret sharing (QSS)~\cite{hillery1999qss,cleve1999how,wei2013experimental,Gu2021differential,jia2021differential} is also important in constructing a secure quantum network with network applications ranging from secure money transfer to multiparty quantum computation.
	
	Secret sharing is a key cryptographic primitive underlying a secure network.
	Secret sharing was first conceived independently by Blakely~\cite{blakley1979safeguarding} and Shamir~\cite{shamir1979how}. 
	It takes both the reliability and secrecy of information into account with practical applications ranging from the management of cryptographic keys, decentralized voting, to a component for secure multiparty computation. 
	In secret sharing, a designated party, called the dealer, divides the secret into shares and distributes them to each player in a way that only authorized subsets of players can reconstruct the secret while all other subsets gain nothing whatsoever.
	The dealer can select a threshold size for authorized subsets.
	For instance, in an $(n,k)-threshold$ scheme, any $k$ $(k\le n)$ of $n$ players can collaborate to recover the secret, while any subset with less than $k$ players remains ignorant.
	
	Classical secret sharing is vulnerable and no longer secure in the face of eavesdroppers equipped with quantum computers.
 
	Fortunately, such threats can be overcome by resorting to quantum technology.
	One can apply quantum key distribution links sharing secure keys between two legitimate users~\cite{bennett1984proceedings,ekert1991quantum,lo2012measurement,braustein2012side,lucamarini2018overcoming,wei2020high,liu2021homodyne,xie2022breaking,zeng2022mode} to establish point-to-point secret keys, which restricts the efficiency in a fully connected quantum network.
	Alternatively, multipartite entangled states---particularly the Greenberger-Horne-Zeilinger (GHZ) entangled states~\cite{greenberger1989bell,mermin1990extreme}---can be used to realize QSS for achieving an advantage over the repetitive use of quantum key distribution links ~\cite{walk2020sharing}. 
	The first QSS protocol was proposed by Hillery \etal using GHZ state for three participants~\cite{hillery1999qss}. This QSS protocol is not secure in the face of participant attacks~\cite{qin2007cryptanalysis}.
	After this protocol, progresses in QSS with multipartite entanglement have been made both in protocols~\cite{li2004efficient,markham2008graph,kogias2017unconditional} and experiments~\cite{chen2005experimental,gaertner2007experimental,peng2018qss} in the past two decades.
	The problem is directly preparing and distributing multipartite states are challenging in practice and limit key rates and transmission distance.
	Therefore, the protocol to distribute postselected GHZ entanglement was proposed to avoid the requirement of entanglement preparation beforehand~\cite{fu2015long}. 
	Although the measurement-device-independent (MDI) protocol needs no entanglement resource, with the increasing number of users, the protocol is limited since the efficiency decays exponentially.
	In addition, the security of QSS protocol in~\cite{fu2015long} is not completely analyzed due to the ignorance of participant attacks.
    To conclude, currently most QSS protocols suffer from decaying transmission efficiency and incomplete security analysis, and thus they are still unpractical for large-scale deployment and application.
	
	To fill the gap of existing protocols, we propose an efficient and practical MDI-QSS protocol based on MDI quantum communication protocols~\cite{lo2012measurement,braustein2012side,fu2015long} and spatial multiplexing and adaptive operation used in all-photonic quantum repeater~\cite{azuma2015all} and adaptive MDI quantum key distribution~\cite{azuma2015allqkd}.
    The results show that our QSS protocol enhances the key rate as the twin-field quantum key distribution does~\cite{wang2018twin,lucamarini2018overcoming,yu2019sending}.
    In terms of security, our protocol is immune to all detection-side attacks which is important for practical quantum communication~\cite{wang2013three,zhou2016making}.
    To be specific, the transmission efficiency of our protocol remains unchanged when the number of communication parties increases.
	Our QSS protocol can break rate-distance bounds~\cite{pirandola2017fundamental} over network under at least ten communication parties when equipped with the GHZ analyzer composed of linear optical elements~\cite{pan1998greenberger}.
	Compared with other protocols, our work improves the secret key rate by more than two orders of magnitude and has a longer transmission distance within an experimentally feasible parameter regime.
	On the other hand, we analyze the security of our protocol in the composable framework considering participant attacks. 
	Based on the security analysis, we also evaluate the performance of our protocol in the finite-size regime.
	Furthermore, we explore applying our QSS protocol as a subroutine to digital signatures, which is a vital primitive in protecting the integrity of data against forgery.
	The digital signatures with our MDI-QSS outperform other quantum counterparts of digital signatures with more than $10^7$ times enhancement in signature rate.
	We believe our protocol manifests the potential to be an important building block for quantum networks.

	\section{Quantum Secret Sharing Protocol}\label{quantum_com}

    \begin{figure}[tbp!]
		\includegraphics[width=8.5cm]{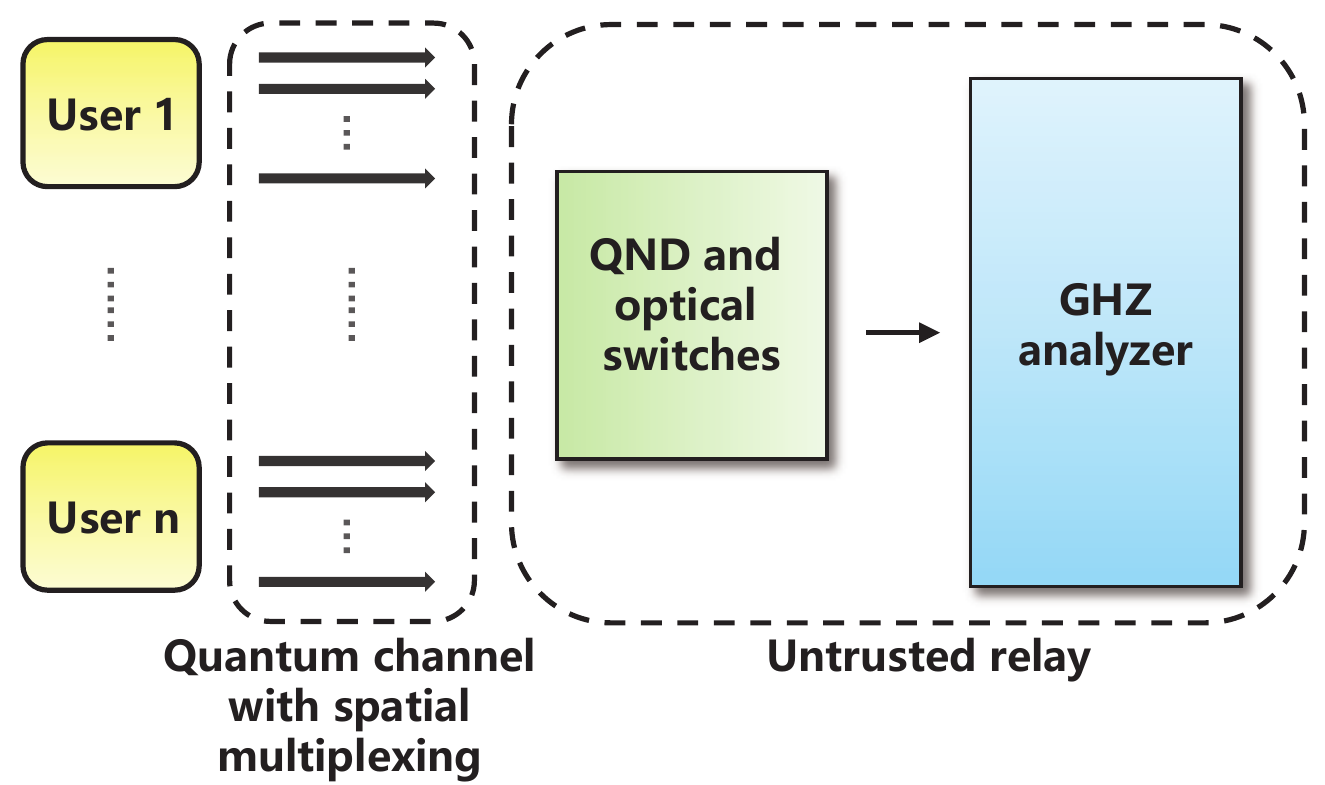}
		\caption{Schematic diagram of our QSS protocol. In our protocol, each user generates $M$ single-photon states selected from eigenstates of the $Z$ and $X$ basis randomly and transmits all $M$ states to the untrusted central relay through the quantum channel with spatial multiplexing. The untrusted central relay performs QND measurements to confirm the arrival of single-photon states. The confirmed photons are routed to the GHZ analyzer via optical switches and the GHZ projection is performed. Each user keeps the information of states that are successfully projected onto the GHZ state and performs classical postprocessing.
		}\label{qss-protocol}
	\end{figure}
 
	Here we consider an $n$-party QSS protocol where the $i$th user is denoted by $A_i$ $(i=1,...,n)$.
	We designate $A_1$ as the dealer dividing and distributing the secret among $n-1$ players ($A_2,...,A_{n}$) and consider an ($n-1,n-1$)-threshold QSS protocol.
    The schematic diagram of our QSS protocol is shown in Fig.~\ref{qss-protocol}.
	
	Before transmitting quantum signals, the dealer $A_1$ establishes a bipartite key with each player to authenticate the classical channel and a joint key as a seed for privacy amplification.
	
	\begin{itemize}
	    \item[$(i)$] 
	    Each user generates $M$ single-photon states that are randomly selected from eigenstates of the $Z$ and $X$ basis.
	    For instance, one selects from $\{\ket{H},\ket{V},(\ket{H}+\ket{V})/\sqrt{2},(\ket{H}-\ket{V})/\sqrt{2}\}$ when using polarization encoding. 
    	He then transmits the $M$ single-photon states to the central relay simultaneously using spatial multiplexing.
        The spatial multiplexing can be realized by using techniques in fiber optical communication like multi-core fiber, multi-mode fiber, mode-division multiplexing, and fiber bundles.
	    
	    \item[$(ii)$]
	    The central relay performs QND measurements to confirm the arrival of single-photon states from $(A_1,... ,A_{n})$.
	    
	    \item[$(iii)$]
	    After the QND measurements, the confirmed photons from every user form a group and are routed to the GHZ analyzer via optical switches.
	    The central relay then performs GHZ projection measurement on the group. 
	    Each user should successfully transmit at least one single photon through QND measurements. 
	    Otherwise, this trial is considered to be failed.
	    
	    \item[$(iv)$]
	    The central relay announces the group information and the GHZ projection results.
	    Each $A_i$ keeps information of states that are successfully projected onto the GHZ state and discards the rest.
	    
	    \item[$(v)$]
	    All $n-1$ players ($A_2,... ,A_{n}$) announce their preparing bases for the remaining trials in any order.
	    If the preparing bases of all $n-1$ players or any single player corresponding to the complementary subset of the remaining $n-2$ players are consistent with the dealer's choice, this round is kept.	    
	    
	    \item[$(vi)$]
        The process is repeated until $m$ rounds in the $X$ basis have been kept for key generation and $k$ rounds in the $Z$ basis have been kept for parameter estimation.
	    Then the dealer calculates the correlation between himself and each single player.
	    If the correlations are below a certain level, the protocol aborts. 
	    
	    \item[$(vii)$]
	    If the correlation test passes, the dealer obtains the raw key and proceeds with error correction leaking a maximum of leak$_{\text{EC}}$ bits of information.
	    To verify the correctness, all $n$ parties compute and compare a hash of length $\log_2(1/\epsilon_{c})$ bits by applying a random universal$_2$ hash function to the raw keys.
	    The protocol aborts if the hash of $A_1$ does not coincide with that of $n-1$ players.
	    If the error correction passes, the dealer conducts privacy amplification using universal$_2$ hashing and obtains the final keys.
	\end{itemize}

	\section{Security analysis}\label{secure_ana}
	
	The security analysis of QSS is quite complex due to the existence of inner malicious parties exploiting the order of announcing the measurement bases and outcomes~\cite{walk2020sharing}.
	The original QSS protocol~\cite{hillery1999qss} consider this problem and can be completely broken~\cite{karlsson1999quantum,qin2007cryptanalysis}.
	In Ref.~\cite{karlsson1999quantum}, the dishonest player (say Charlie) intercepts all the GHZ photons from the dealer and establishes Bell entanglement between himself and the other player. 
	Once Charlie obtains the knowledge of other players' measurement bases, he can learn their measurement outcomes as well through Bell entanglement.
	Furthermore, Charlie can ensure the round will be kept if the dealer chooses the same basis as him and recreates the dealer's information.
	As a result, the whole protocol is broken while Charlie remains undetected.
	Qin \etal provided a general result of the necessary and sufficient conditions under which Charlie can attain all the information without being detected~\cite{qin2007cryptanalysis}.
	
	To address the participant attacks, Kogias \etal proposed to treat the measurements announced by the players as an input or output of an uncharacterized measuring device and the dealer as a trusted party with trusted devices.
	Then the security of QSS can be connected with one-sided device-independent quantum key distribution which has been proven unconditionally secure~\cite{kogias2017unconditional}.
	Similarly, Refs.~\cite{williams2019quantum,grice2019quantum,gu2021secure,shen2023experimental} applied the security proof of standard quantum key distribution with trusted devices in both discrete and continuous variable QSS.
	Walk \etal stated the essential part of the security proof in Ref.~\cite{kogias2017unconditional} was excluding the potential malicious parties from parameter estimation~\cite{walk2020sharing}.
	As a comparison, in Ref.~\cite{williams2019quantum}, the dealer randomly selects a set of potential malicious parties and includes them in parameter estimation.
	However, the potential malicious parties are forced to make announcements first.
	In our QSS protocol, we follow Refs.~\cite{kogias2017unconditional,walk2020sharing} as shown in $(v)$ and $(vi)$ of our protocol to prevent dishonest participants.
	
	We introduce some useful definitions in the following description.
	In general, the dealer's final key \textbf{S} can be quantum mechanically correlated with a quantum state held by the adversary, and such a state is described by the classical-quantum state
	\begin{equation}
		\rho_{\textbf{S},EU_j}=\sum_{\textbf{S}}p(\textbf{S})\ket{\textbf{S}}\bra{\textbf{S}}\otimes\rho^{\textbf{S}}_{E,U_j},
	\end{equation}
	where the sum is over all possible strings and $\rho^{\textbf{S}}_{E,U_j}$ is the joint state of the eavesdropper and the $j$th untrusted subset given \textbf{S}.
    In our work, we consider a general adversary which is able to perform any operations permitted by quantum laws rather than a specific adversary model performing concrete attacks.
	By untrusted subset, we mean the subset formed by any $n-2$ players.
	Thus we have $n-1$ untrusted subsets in total in our QSS protocol.
    $p(\textbf{S})$ is the probability of the state $\ket{\textbf{S}}\bra{\textbf{S}}\otimes\rho^{\textbf{S}}_{E,U_j}$ presenting in the $\rho_{\textbf{S},EU_j}$.
	Ideally, a QSS protocol is secure if it is correct and secret.
	The correctness means the dealer's bit strings \textbf{S} are identical to the bit strings $\textbf{S}_{\text{player}}$ recreated from all $n-1$ players, $i.$ $e.$ $\textbf{S}=\textbf{S}_{\text{player}}$.
	The secrecy requires $\rho_{\textbf{S},EU_j}=\sum_{\textbf{S}}\frac{1}{|\textbf{S}|}\ket{\textbf{S}}\bra{\textbf{S}}\otimes\sigma_{EU_j}$, which means the joint system of the eavesdropper and the $j$th untrusted subset is decoupled from the dealer.
	However, these two conditions can never be met perfectly.
	In practice, we call a QSS protocol $\epsilon_c$-correct if
	\begin{equation}\label{def_cor}
		\text{Pr}\left( \textbf{S}\neq\textbf{S}_{\text{player}}\right) \le\epsilon_c.
	\end{equation}
	We call a QSS protocol $\epsilon_s$-secret if
	\begin{equation}\label{def_sec}
		\max_j\left\lbrace p_{\text{pass}}D\left( \rho_{\textbf{S},EU_j},\sum_{\textbf{S}}\frac{1}{|\textbf{S}|}\ket{\textbf{S}}\bra{\textbf{S}}\otimes\sigma_{EU_j}\right) \right\rbrace \le\epsilon_s,
	\end{equation}
	where $D(\cdot,\cdot)$ is the trace distance and $p_{\text{pass}}$ is the probability that the protocol does not abort.
	The maximization is over all $n-1$ untrusted subsets since the dealer must take worst-case estimates for the secrecy.
	A QSS protocol is called $\epsilon_{sec}$-secure with $\epsilon_{sec}\ge\epsilon_{s}+\epsilon_{c}$ if it is $\epsilon_c$-correct and $\epsilon_s$-secret.
	
	Similar to quantum key distribution~\cite{tomamichel2012tight}, the extractable amount of key $l$ for a $\epsilon_c$-correct and $\epsilon_s$-secret QSS is
	\begin{equation}
		l=\min_jH^{\epsilon}_{\text{min}}(\textbf{X}|EU_j)-\text{leak}_\text{EC}-\log_2\frac{1}{\epsilon_c\bar{\epsilon}^2}+2,
	\end{equation} 
	where $H^{\epsilon}_{\text{min}}(\textbf{X}|EU_j)$ is the conditional smooth min-entropy characterizing the average probability that the eavesdropper and dishonest parties guess the dealer's raw key \textbf{X} correctly using optimal strategy and leak$_{\text{EC}}$ is the amount of information leakage of error correction.
	$\epsilon$ and $\bar{\epsilon}$ are positive constants proportional to $\epsilon_{s}$.
	For a realistic scenario, the computable key length of QSS is
	\begin{equation}\label{QSSlength}
            \begin{split}
                l=m&\left[ q-\max_jh( E_Z^{AA_j}+\mu( E_Z^{AA_j},\epsilon'))\right]\\
                &-\text{leak}_\text{EC}-\log_2\frac{4}{\epsilon_c\bar{\epsilon}^2},
            \end{split}	
	\end{equation}
	where $\mu(\lambda,\epsilon)=\frac{\frac{(1-2\lambda)AG}{m+k_j}+\sqrt{\frac{A^2G^2}{(m+k_j)^2}+4\lambda(1-\lambda)G}}{2+2\frac{A^2G}{(m+k_j)^2}}$,	with $k_j(<k)$ being the number of parameter estimation rounds between the dealer and the complementary single player of the $j$th untrusted subset, $\lambda$ being the error rate observed in parameter estimation, $A=\max\{m,k_j\}$, and $G=\frac{m+k_j}{mk_j}\ln\frac{m+k_j}{2\pi mk_j\lambda(1-\lambda)\epsilon^2}$.
	$E_Z^{AA_j}$ is the marginal error of the correlation test.
    $q$ is a constant that quantifies the complementary of the two preparing bases.
	We give a full proof and analysis of the extractable key length in Appendix~\ref{securityproof}.
		
	\section{Performance}\label{performance_qcom}
	In this section, we evaluate the performance of our QSS protocol.
	We introduce a benchmark used in our investigation and analyze the performance of our protocol under both the asymptotic and finite-size regime.
	In the end, we utilize our QSS as a key generation solution to an essential cryptographic primitive---digital signatures and investigate the signature rate of signing a document.
	
	\subsection{Asymptotic performance of MDI-QSS}\label{performance}
	
	In the asymptotic limit, we follow the key rate formula presented in~\cite{fu2015long}.
	To be specific, Fu \etal proposed the secret key rate of MDI-QSS for the first time~\cite{fu2015long,lo2012measurement,braustein2012side,maneva2002improved,dur1999separability,bennett1996mixed,gottesman2004security,azuma2015allqkd}
	\begin{equation}\label{qssrate}
		R_{\text{QSS}}=Q_X\left[ 1-h(E_Z)-fh(E_X)\right] ,
	\end{equation}
	where $Q_X$ is the gain of the $X$ basis, the probability of successful GHZ state projection when preparing a single photon in the $X$ basis, and $E_X$ ($E_Z$) is the bit (phase) error rate.
	$h(x)=-x\log_2x-(1-x)\log_2(1-x)$ is the binary Shannon entropy function. $f$ is the inefficiency of error correction.
	
	The gain $Q_X$ is defined as the efficiency of successfully generating postselected GHZ entanglement when preparing a single photon in the $X$ basis.
	Specifically, we have $Q_{X}=\frac{\bar{N}}{M}$, where $\bar{N}$ is the average number of successful GHZ projection formed by photons using $M$ multiplexing.
	If we denote the total efficiency of both GHZ projection and the channel from any $i$th user to the central node as $\eta_{\rm tot}$, and $M$ multiplexing is used, then $\bar{N}\sim M\eta_{\rm tot}$.
	Therefore, we have $Q_X\sim\eta_{\rm tot}$.
	The approximate relation can be converted to an equation $Q_X=\eta_{\rm tot}$ under the asymptotic limit ($M\rightarrow\infty$). 
	We prove this equation when $n=3$ in Appendix~\ref{derivation}.
	To guarantee that more than one entanglement is generated on average, the multiplexing number should satisfy $M\ge \eta_{\rm tot}^{-1}$, which implies that $\bar{N}\sim M\eta_{\rm tot}\ge1$.
	
	In this simulation, we use efficiency $\eta_{\text{sps}}$ to describe the probability of the single photon source generating single photons and set $\eta_{\text{sps}}=0.9$~\cite{christensen2013detection}.
    We consider the GHZ analyzer based on linear optical elements~\cite{pan1998greenberger} capable of identifying two of the $n$-particle GHZ states.
	We present the detailed working of the analyzer in Appendix~\ref{GHZanalyzer}.
	Photons travel through optical fiber channels whose transmittance is determined by $\sqrt{\eta_{\text{channel}}}=\exp\left(-\frac{l}{l_{\text{att}}} \right)$, where the attenuation distance $l_{\text{att}}=27.14$ km and $l$ is the distance from any $i$th user to the GHZ analyzer.
	QND measurements are required to confirm the arrival of photons and the success probability of QND measurements is denoted by $p_{\text{QND}}$.
	To simplify the simulation, we consider a QND measurement for a single photon based on quantum teleportation~\cite{kok2002single} with ideal parameters where we have $p_{\text{QND}}=1/2$. 
	The active feedforward technique is needed to direct the arrived photons to the GHZ analyzer via optical switches.
	We assume the active feedforward costs time $\tau_a=67$ ns~\cite{ma2011experimental}, which is equivalent to a lossy channel with the transmittance $\eta_a=\exp(-\tau_ac/l_{\text{att}})$, where $c=2.0\times10^8$ $\text{ms}^{-1}$ is the speed of light in an optical fiber.
	Single photon detectors in the GHZ analyzer are characterized by an efficiency of $\eta_d=0.93$ and a dark count rate of $p_d=1\times10^{-9}$~\cite{minder2019experimental}, by which we can estimate the success probability of GHZ projection in the $X(Z)$ basis $Q_{X(Z)}^{\text{GHZ}}$.
	Based on the aforementioned assumption on experiment parameters, we analytically estimate the gain with
	\begin{equation}
		Q_{X}=Q^{\text{GHZ}}_{X}\cdot p_{\text{QND}}\cdot \sqrt{\eta_{\text{channel}}}\cdot \eta_{\text{sps}}\cdot \eta_a.
	\end{equation}
	See Appendix~\ref{estimation} for the concrete process of estimation of the marginal bit error rates and phase error rate.  
	
	\begin{figure}[tbp!]
		\includegraphics[width=8.5cm]{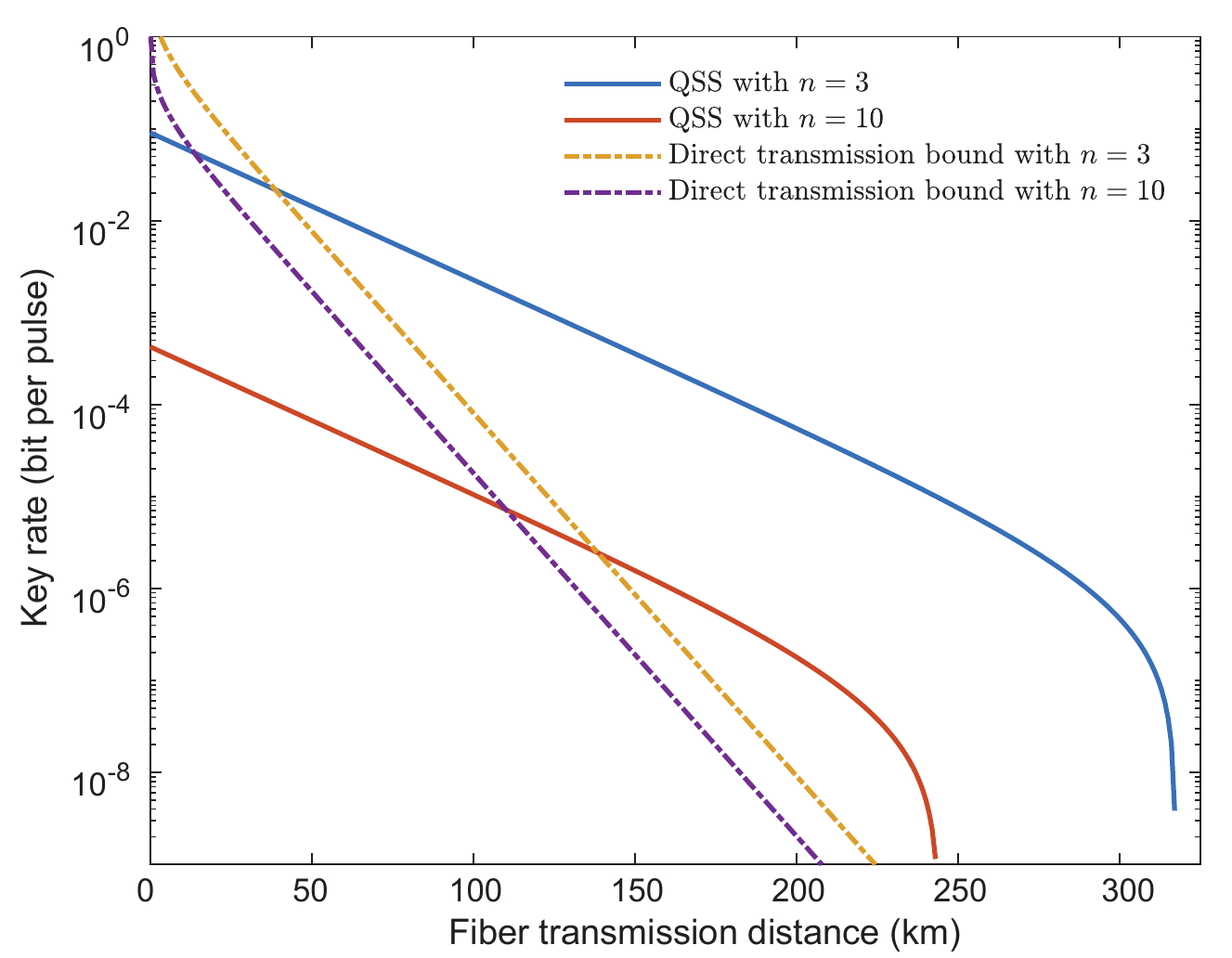}
		\caption{Key rates of our QSS and direct transmission bounds. We show key rates of our protocol and corresponding bounds under different numbers of communication parties ($n=3, 10$ from top to bottom). In the figure, key rates of our protocol and bounds are plotted with solid and dash-dotted lines, respectively. The fiber transmission distance denotes the distance between any $i$th party and the central relay. 
		}\label{qss_vs_bound}
	\end{figure}
	
	Before analyzing the performance of our protocols, we discuss the limitations on quantum communication over network and provide a benchmark for our protocol.

	\textcolor{black}{A general methodology allowing to upperbound the two-way capacities of an arbitrary quantum channel with a computable single-letter quantity was devised in~\cite{pirandola2017fundamental}, which determines the fundamental rate-loss tradeoff affecting any quantum key distribution protocol.
	In this way, for the lossy channel, they proved that the two-way quantum capacity and the secret-key capacity are $-\log_2(1-\eta)$, which is the maximum rate achievable by any optical implementation of point-to-point quantum key distribution.
    This bound sets the limits of point-to-point quantum communications and provides precise and general benchmarks for quantum repeaters.}
    For quantum communications over network scenarios, bounds have also been established under different scenarios~\cite{pirandola2019end,pirandola2020general}.
    \textcolor{black}{In~\cite{laurenza2017general}, the methodology used in~\cite{pirandola2017fundamental} is extended to a more complex communication scenario including quantum broadcast channel, quantum multiple-access channel, and all-in-all quantum communication, where multiple senders and/or receivers are involved.}
	Later, Das \etal provided a unifying framework to upperbound the key rates of both bipartite and conference settings with different scenarios including broadcast, multiple access, interference channels, and more general network scenarios~\cite{das2021universal}.

	\begin{figure}[tbp!]
		\includegraphics[width=8.5cm]{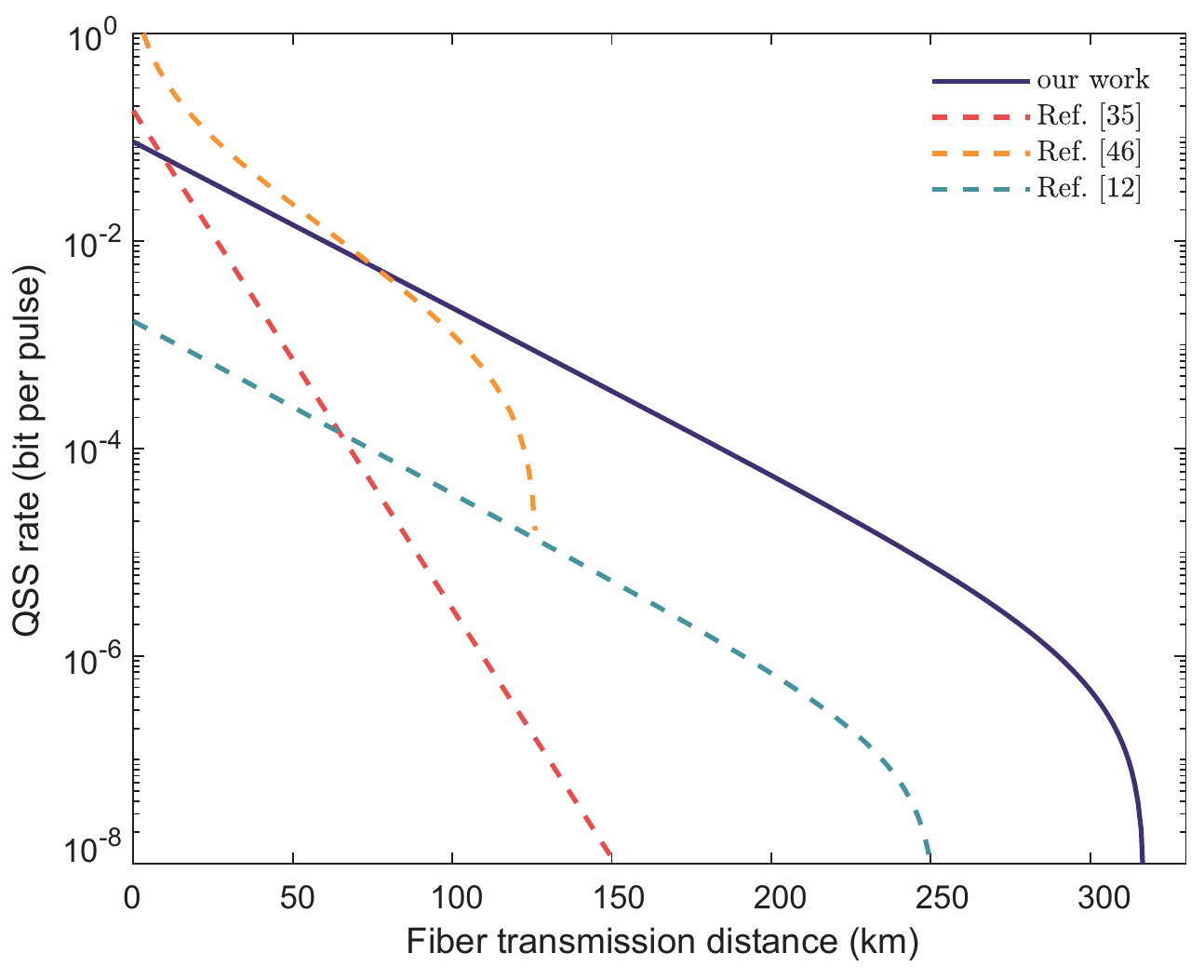}
		\caption{Comparison of key rates of QSS from our work, original MDI-QSS~\cite{fu2015long}, continuous variable (CV) QSS~\cite{grice2019quantum}, and twin-field (TF) differential phase shifting (DPS) QSS~\cite{Gu2021differential}. We plot the key rates of the protocols when $n=3$. Different colored lines are used to denote different protocols. The fiber transmission distance denotes the distance between any $i$th party and the central relay.
		}\label{figqss}
	\end{figure}

    In our work, to investigate the performance of our protocol, we consider a rate benchmark in a case where the untrusted central node is removed and all $n$ users are linked by a star network similar to that in Ref.~\cite{Grasselli2019conference}. 
    In such a scenario, a selected user performs quantum key distribution with other users $n-1$ times to establish bipartite secret keys with the same length due to the network symmetry.
    According to the secret-key capacity, the asymptotic rate is $-\log_2(1-\eta)$ with $\sqrt{\eta}$ being the transmittance between any $i$th user and the central node.
    The selected user can XOR all $n-1$ key strings to conduct secret sharing.
    The final key length is equal to the keys' lengths obtained using quantum key distribution.
    Therefore, in this scenario, the key rate is bounded by $\frac{-\log_2(1-\eta)}{n-1}$.
    We call this bound the direct transmission bound.
    It should be noted that the above scenario does not necessarily yield the highest key rate in secret sharing.
	
	In Fig.~\ref{qss_vs_bound}, we plot the key rates of our QSS as well as direct transmission bounds with different numbers of communication parties.
	We present key rates and bounds with $n=3,10$ users from top to bottom using solid and dash-dotted lines respectively.
    Our protocol breaks the direct transmission bounds because of the spatial multiplexing and adaptive operations. 
	A polynomial scaling of efficiency with distance can be realized for at least ten users over the network while the bounds attenuate greatly as $n$ increases.

	To further investigate the performance of our work, we evaluate the key rate of our protocol and that of other preceding QSS protocols over a quantum network under the same experimental parameters.
	In Fig.~\ref{figqss}, we plot the key rate of our QSS protocol, original MDI-QSS~\cite{fu2015long}, continuous variable (CV) QSS~\cite{grice2019quantum}, and twin-field  (TF) differential phase shifting (DPS) QSS~\cite{Gu2021differential} with $n=3$.
	We can directly conclude from Fig.~\ref{figqss} our work can achieve a longer transmission distance of more than 300 km and increase the secret key rate by at least two orders of magnitude at long distances compared with other QSS protocols.
	Though TF DPS QSS achieves a similar transmission distance and slope to our work, the TF DPS QSS protocol only works with three communication users and cannot be easily and directly extended to scenarios when $n$ is more than three.
    The CV QSS protocol can reach no more than 140 km.
	One can observe that CV QSS outperforms our work at shorter distances because CV protocols adopt the coherent state as information carrier which is more robust to channel loss.
	As a result, the signals can always be detected, which means the gain of CV protocol is always unity. 
	The CV QSS protocol is asymmetric where the dealer measures the Gaussian signals from the users while our QSS is symmetric in the quantum phase of the protocol.
	Therefore, the CV QSS is not as flexible as our QSS to deploy in the quantum network.

	\subsection{Performance of QSS in finite-size regime}

	We investigate the performance of our QSS protocol in the finite-size regime with the same parameters introduced in the asymptotic scenario.
	Wse fix $\epsilon_{c}=10^{-15}$ corresponding to a realistic hash tag size in practice~\cite{renner2008security}.
	In our QSS protocol, for simplicity, we assume the information leakage during error correction to be $\text{leak}_{\text{EC}}=fh(E_X)$, where $f=1.1$, $h(x)$ is the binary Shannon entropy, and $E_X$ is the error rate in the $X$ basis.
	Then following Eq.~(\ref{QSSlength}) we can obtain the result in finite-size regime.

	\begin{figure}[tbp!]
	\includegraphics[width=8.5cm]{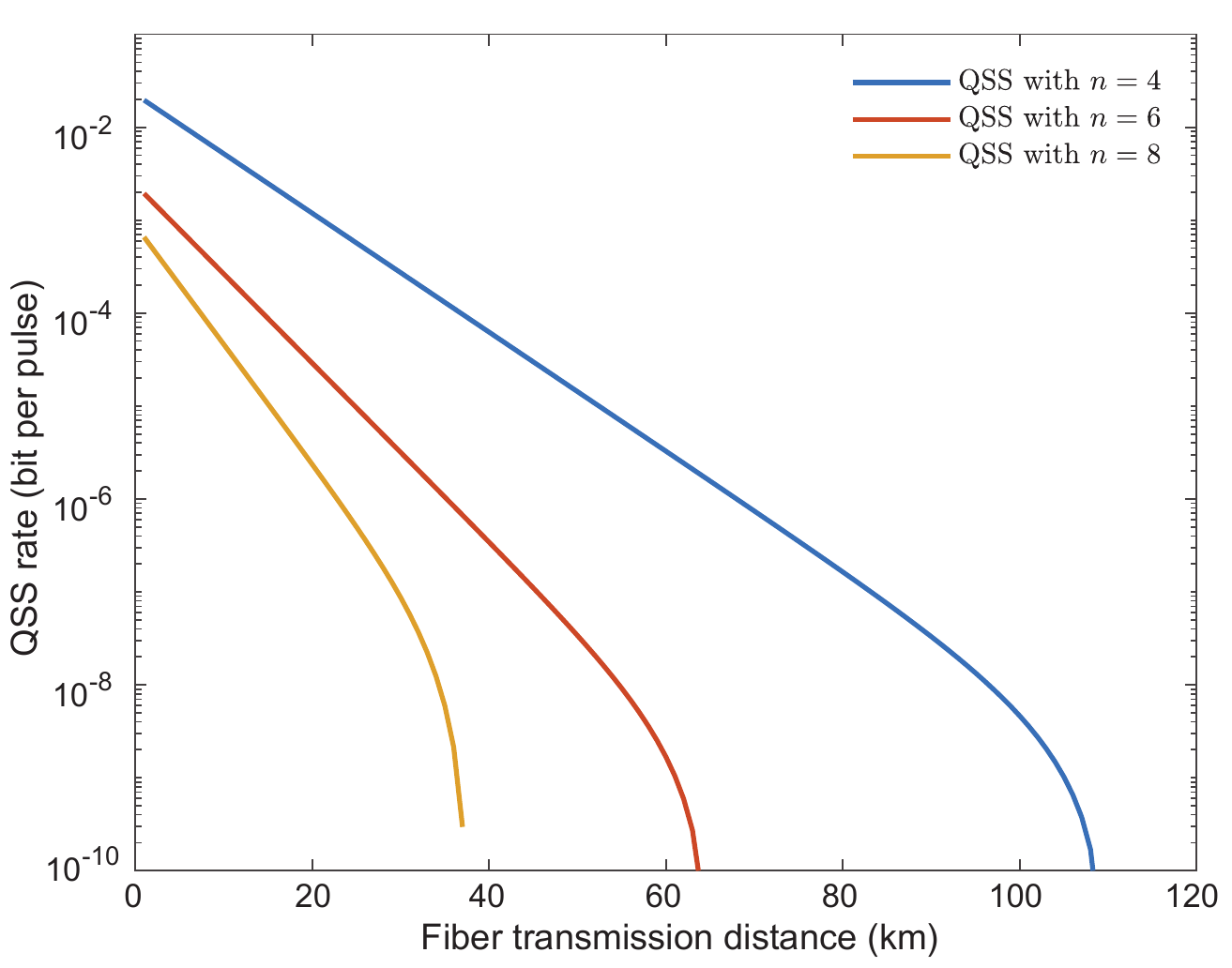}
		\caption{Secret key rate of our QSS as a function of distance in finite-size regime. We consider the secret key rate of QSS with $n=4,6,8$ shown in different colors. In this simulation, we fix the total number of signals to be $10^{12}$. The fiber transmission distance denotes the distance between any $i$th party and the central relay. 
		}\label{qss_dis_finite}
	\end{figure}
 
	In Fig.~\ref{qss_dis_finite}, we plot the secret key rate of our QSS protocol as a function of the distance between any $i$th user and the central relay.
	We can view that our QSS can transmit more than 100 km, 60 km, and 30 km when $n=4,6,8$, respectively.
	These transmission distances can cover the intra- and inter-city deployment of the quantum network.
    On the other hand, with the all-photonic nature of our QSS protocol, our work is feasible and can be implemented with state-of-the-art technology.
    Combining these two factors, our results are meaningful to the practical deployment of a quantum network.
	The slope of the curve is observed to differ with different values of $n$, which stems from the secret key rate here counts the probability of all users choosing the same basis which scales exponentially with $n$.
	
	In the above two subsections, we investigate our protocol under a model consisting of single photon sources, QND measurements, optical switches, and the GHZ analyzer based on linear optical elements. 
	Our protocol can be improved with other techniques.
	For instance, our protocol can be improved by utilizing the complete GHZ analyzer which can identify all $2^n$ GHZ states, such as GHZ state analysis taking into account nonlinear processes~\cite{qian2005universal,Xia2014complete} or entangled-state analysis for hyperentangled photon pairs~\cite{sheng2010complete,liu2015generation}.
	On the other hand, in step $(iii)$ from Sec.~\ref{quantum_com}, large-scale optical switches are needed to route the photons into the GHZ analyzer, which may affect the transmittance and cause unwanted loss.
	Thus, future effort should be made towards realizing the protocol with reduced scale optical switches and one possible way is utilizing a Hadamard linear optical circuit together with single-mode on/off switches~\cite{azuma2015allqkd}.
	Techniques in MDI quantum key distribution~\cite{zhou2016making,GU2022experimentalMDI} can be applied in our QSS to further improve practicality.

	\subsection{Key Generation Solution For Quantum Digital Signatures}
	
	Digital signatures, as an important cryptographic primitive, promise the authenticity, integrity, and non-repudiation of information processing, which have been applied in various areas such as financial transactions, software distribution, and blockchain. 
	The security of classical digital signatures is based on the complexity of mathematical problems.
	While the quantum counterpart of digital signatures, called quantum digital signatures (QDSs), guarantees security via the laws of quantum physics.
	Since the first QDS protocol which is challenging in the experiment, progresses have been made to improve the practicality of QDS~\cite{yin2016practical,amiri2016secure,Lu2021efficient}.
	However, the existing protocols suffer from low signature rate and are unpractical when signing multi-bit documents.
	Yin \etal proposed a QDS protocol capable of signing long documents with information-theoretic unconditional security~\cite{yin2022experimental}.
	The QDS protocol builds a perfect bit correlation of three users with an asymmetric key system and realizes an efficient QDS together with completely random universal$_2$ hash function and one-time pad.
	Our QSS is capable of generating perfect key correlations between any $n$ users, which naturally fits well in the framework of such QDS protocol.
	Furthermore, our protocol has great potential and capability of large-scale application of such QDS in the future quantum network. 
	Thus here we investigate the performance of applying our QSS protocol as a subroutine in the key distribution process of~\cite{yin2022experimental}.
	
	We start with briefly introducing this QDS protocol.
	For convention, let Alice be the signer with Bob and Charlie as the receiver.
	Before generating and verifying digital signatures, perfect key correlations $X_A=X_B\oplus X_C$ $(Y_A=Y_B\oplus Y_C)$ should be realized among Alice, Bob, and Charlie, where $X_i$ $(Y_i)$ $(i=A,B,C)$ denotes secret keys held by each user.
	QSS can achieve such correlations and thus our QSS protocol provides a natural solution to the key generation process.
	After obtaining the keys, Alice generates digital signatures of an arbitrary document through completely random universal$_2$ hash function and one-time pad and transfers the signed document to Bob.
	Bob transmits his key bit strings and the signed document to Charlie.
	Bob and Charlie verify the digital signatures and if both of them accept the signed document we can say this is a successful signing.
	For more technical details, Ref.~\cite{yin2022experimental} can be referred to.
	
	We investigate the performance of QDS protocol in \cite{yin2022experimental} using our QSS to generate perfect key correlations.
	It is further compared with the experiment result of QDS protocol with quantum states exchanged forward in~\cite{richter2021agile}, which is shown in Tab. \ref{tab2}.
	For the calculation of QDS using our QSS, we assume the order of the irreducible polynomial to be 128 which indicates a security bound about $10^{-34}$~\cite{yin2022experimental} and set the system clock frequency to be 1 MHz.
	In order to have a direct comparison between the two protocols, in Tab. \ref{tab2} we calculate and list the signature rate of signing a document with the size of $10^6$ bits, which indicates the amount of documents signed per second.
	From the comparison, we can easily conclude that the QDS protocol with keys generated by our QSS outperforms the QDS in~\cite{richter2021agile} with a better signature rate and longer distance.
	Our QSS shows great practicality when used in QDS protocol.

	\begin{table}[t!]
		\caption{Performance of QDS protocol using our QSS and QDS with quantum states exchanged forward in~\cite{richter2021agile}. The performance of the QDS protocols is evaluated by the signature rate of signing a document with the size of $10^6$ bits. We assume the system clock frequency to be 1 MHz. NaN means no digital signatures can be generated. The unit of signature rate is times per second (tps).}
		\begin{tabular}{ccc}
			\hline
			\hline
			&Distance (km)&Signature rate (tps)\\
			\hline
			\multirow{2}{*}{QDS~\cite{yin2022experimental} with our QSS}&20&162\\
			&50&93\\
			\multirow{2}{*}{QDS in~\cite{richter2021agile}}&20&7.3$\times10^{-6}$\\
			&50&NaN\\
			\hline
			\hline
		\end{tabular}	\label{tab2}
	\end{table}
	
	\section{Conclusion and outlook}\label{conclusion}
	
	In this work, we propose an MDI-QSS protocol for quantum network applications. 
        Our QSS can break the rate-distance bound with the GHZ analyzer based on linear optical elements under at least ten network users.
	By comparing our work with the key rate of recent QSS works, we show the superiority of our work by improving the key rate by more than two orders of magnitude and achieving longer transmission distances.
	The security of our QSS taking the participant attacks into account is analyzed in the composably secure framework.
	Based on the security analysis, we provide a computable key length in the finite-size regime. 
	Furthermore, we consider applying QSS to another important cryptographic primitive--QDS.
	The result shows that QDS with our MDI-QSS protocol as a subroutine possesses significantly higher efficiency compared with preceding QDS.
	Based on the result of this work, we can anticipate a wide and flexible usage of our work in multiparty applications of the secure quantum network.
	
	Here we remark on possible directions for future work.
	In conventional quantum repeater protocols~\cite{duan2001long,simon2007quantum,fault2006childress,hybrid2006loock}, quantum memories are necessary to be entangled with photons and to preserve entanglement at least until receiving heralding signals of successful entanglement swapping.
	Here time multiplexing from quantum memories' preserving entanglement enables the enhancement in transmission efficiency.  
	On the other hand, all-photonic quantum repeater protocol~\cite{azuma2015all}, requiring no matter qubit quantum memories and demonstrating polynomial scaling of efficiency with distance, was proposed. 
	The all-photonic scheme utilizes cluster states to realize a polynomial scaling with distance which is in fact a result of spatial multiplexing. 
	Therefore, with such spatial multiplexing idea, we can develop other protocols apart from quantum communication with enhanced efficiency.
	On the other hand, secret sharing can be useful in constructing protocols such as Byzantine consensus and federated learning.
	Our work can be applied to these protocols as a subroutine for improved efficiency and security against eavesdroppers with quantum computer.
	In addition, our work can be further developed to give anonymity to users~\cite{grasselli2022anonymous} over quantum network for more complex application scenarios.
	
	\section*{Acknowledgments}
	We gratefully acknowledge the supports from the National Natural Science Foundation of China (No. 12274223), the Natural Science Foundation of Jiangsu Province (No. BK20211145), the Fundamental Research Funds for the Central Universities (No. 020414380182), the Key Research and Development Program of Nanjing Jiangbei New Area (No. ZDYD20210101),  the Program for Innovative Talents and Entrepreneurs in Jiangsu (No. JSSCRC2021484), and the Program of Song Shan Laboratory (Included in the management of Major  Science and Technology Program of Henan Province) (No. 221100210800-02).
	
	\appendix
	
	\section{Security proof}
	\label{securityproof}
	In this appendix, we provide detailed process to prove the security of our QSS protocol and show how to get the computable key length Eq.(\ref{QSSlength}).
	
	\subsection{Security proof of QSS}

    As we have introduced in Sec.~\ref{secure_ana}, a QSS protocol is secure if it is correct and secret.
	The correctness means the dealer's bit strings \textbf{S} are identical to the bit strings $\textbf{S}_{\text{player}}$ recreated from all players.
	The secrecy requires the joint system of the eavesdropper and the $j$th untrusted subset is decoupled from the dealer.
	However, these two conditions can never be met perfectly.
	In practice, we call a QSS protocol $\epsilon_c$-correct if it satisfies Eq. (\ref{def_cor}) and $\epsilon_s$-secret if it satisfies Eq. (\ref{def_sec}).
	A QSS protocol is called $\epsilon_{\rm sec}$-secure with $\epsilon_{\rm sec}\ge\epsilon_{s}+\epsilon_{c}$ if it is $\epsilon_c$-correct and $\epsilon_s$-secret.
    Therefore, to prove the $\epsilon_{sec}$-security of a QSS protocol, we should prove the $\epsilon_{c}$-correctness and $\epsilon_{s}$-secrecy of our QSS. 
    In the following, we prove the $\epsilon_{c}$-correctness and $\epsilon_{s}$-secrecy of our QSS in \textbf{Theorem 1} and \textbf{Theorem 3}.
    Based on these two theorems, we can guarantee the $\epsilon_{sec}$-security of our QSS protocol and thus finish the security analysis of our protocol in the composable framework.
    
	\textbf{Theorem 1.} \textit{The QSS protocol defined in Sec.~\ref{quantum_com} is $\epsilon_{c}$-correct.}
	
	\begin{proof}
	In step $(vii)$ of QSS, all $n$ parties compute and compare a hash of length $\log_2(1/\epsilon_{c})$ by applying a random universal$_2$ hash function to raw keys \textbf{X} and \textbf{X}$_{\text{player}}$.
	If the hash value disagrees, the protocol aborts.
	According to the property of universal$_2$ hash function~\cite{cover1991information}, the probability that two hash values coinciding---if \textbf{X} and \textbf{X}$_{\text{player}}$ are different and the hash function is chosen uniformly at random from the family---is at most $2^{\lceil\log_2\epsilon_{c}\rceil}\le\epsilon_{c}$.
	Therefore, it is guaranteed that $\text{Pr}(\textbf{S}\neq\textbf{S}_{\text{player}})\le\text{Pr}(\textbf{X}\neq\textbf{X}_{\text{player}})\le\epsilon_{c}$.
	\end{proof}
	
	To prove that our QSS protocol is $\epsilon_{s}$-secret, we introduce the Quantum Leftover Hashing Lemma~\cite{brus1998optimal}.
	
	\textbf{Lemma 2.} \textit{If Alice uses a random universal$_2$ hash function to map the raw key \textbf{X} to the final key \textbf{S} and extracts a string of length $l$, then for any positive $\epsilon$
	\begin{equation}
		D\left( \rho_{\textbf{S},E},\sum_{\textbf{S}}\frac{1}{|\textbf{S}|}\ket{\textbf{S}}\bra{\textbf{S}}\otimes\sigma_{E}\right) \le\sqrt{2^{l-H^{\epsilon}_{\rm min}(\textbf{X}|E')-2}}+2\epsilon,
	\end{equation}
	where $E$ is a finite or infinite dimensional system of Eve and $E'$ summarizes all information Eve obtained including the classical communication.}
	
	Now we can prove the $\epsilon_{s}$-secrecy of our QSS protocol.
	
	\textbf{Theorem 3.} \textit{The QSS protocol defined in Sec.~\ref{quantum_com} is $\epsilon_{s}$-secret if the key length $l$ satisfies
	\begin{equation}
            \begin{split}
                l=&m\left[q-\max_jh(E_Z^{AA_j}+\mu(E_Z^{AA_j},\epsilon'))\right]\\
                &-\text{\rm leak}_\text{EC}-\log_2\frac{4}{\epsilon_c\bar{\epsilon}^2},
            \end{split}
	\end{equation}
	where
	\begin{equation}
		\mu(\lambda,\epsilon)=\frac{\frac{(1-2\lambda)AG}{m+k_j}+\sqrt{\frac{A^2G^2}{(m+k_j)^2}+4\lambda(1-\lambda)G}}{2+2\frac{A^2G}{(m+k_j)^2}}
	\end{equation}
	with $k_j(<k)$ being the number of parameter estimation rounds between the dealer and the complementary single player of the $j$th untrusted subset, $\lambda$ being the error rate observed in parameter estimation, $A=\max\{m,k_j\}$ and $G=\frac{m+k_j}{mk_j}\ln\frac{m+k_j}{2\pi mk_j\lambda(1-\lambda)\epsilon^2}$.
	$\epsilon$ and $\bar{\epsilon}$ are positive constants proportional to $\epsilon_{s}$.
	}

	\begin{proof}
	To fit in our QSS protocol, the Eve's system in the Quantum Leftover Hashing Lemma includes both eavesdropper and the $j$th untrusted party $U_j$.
	By choosing $\epsilon=(\epsilon_{s}-\bar{\epsilon})/(2p_{\text{pass}})$ with $\bar{\epsilon}>0$ and
	\begin{equation}
		l=H^{\epsilon}_{\text{min}}(\textbf{X}|EU_j')+2-2\log_2\frac{p_{\text{pass}}}{\bar{\epsilon}},
	\end{equation}
	we have
	\begin{equation}
		p_{\text{pass}}D\left( \rho_{\textbf{S},EU_j},\sum_{\textbf{S}}\frac{1}{|\textbf{S}|}\ket{\textbf{S}}\bra{\textbf{S}}\otimes\sigma_{EU_j}\right) \le\epsilon_s.
	\end{equation}
	By taking the maximum over all $j$, we can reach a $\epsilon_{s}$-secret QSS protocol.
	Furthermore, using the fact that $\log_2p_{\text{pass}}<0$, we choose the key length
	\begin{equation}
		l=H^{\epsilon}_{\text{min}}(\textbf{X}|EU_j')+2-2\log_2\frac{1}{\bar{\epsilon}}
	\end{equation}
	to ensure a $\epsilon_{s}$-secret QSS protocol.
	Now we present how to obtain key length in Eq.~(\ref{QSSlength}).
	During error correction the amount of leak$_{\text{EC}}$$+\log_2(1/\epsilon_{c})$ bits of information about the dealer's raw key \textbf{X} are revealed and we have~\cite{tomamichel2012tight}	
	\begin{equation}
		H^{\epsilon}_{\text{min}}(\textbf{X}|EU_j')\ge H^{\epsilon}_{\text{min}}(\textbf{X}|EU_j)-\text{leak}_{\text{EC}}-\log_2\frac{1}{\epsilon_{c}}.
	\end{equation}
	All that remains is to lower bound the conditional smooth min-entropy and this can be achieved by using entropic uncertainty relation~\cite{tomamichel2011uncertainty}
	\begin{equation}\label{EUR}
		H^{\epsilon}_{\text{min}}(\textbf{X}|EU_j)+H^{\epsilon}_{\text{max}}(\textbf{Z}|C_j)\ge mq,
	\end{equation}
	where $C_j$ is the complementary trusted player of untrusted subset $U_j$ and $q$ is the preparation quality quantifying the incompatibility of two measurements~\cite{tomamichel2011uncertainty,tomamichel2012tight}.
	From Eq.~(\ref{EUR}), we can lower bound the conditional smooth min-entropy using the smooth max-entropy $H^{\epsilon}_{\text{max}}(\textbf{Z}|C_j)$ characterizing the correlations between \textbf{Z} and $C_j$.
	There is only one single player in $C_j$ and we can apply the result of quantum key distribution~\cite{yin2020tight}
	\begin{equation}
		H^{\epsilon}_{\text{max}}(\textbf{Z}|C_j)\le mh(E_Z^{AA_j}+\mu(E_Z^{AA_j},\epsilon')),
	\end{equation}
	where $\epsilon=\epsilon'/\sqrt{p_{\text{pass}}}$ and 	\begin{equation}
		\mu(\lambda,\epsilon)=\frac{\frac{(1-2\lambda)AG}{m+k_j}+\sqrt{\frac{A^2G^2}{(m+k_j)^2}+4\lambda(1-\lambda)G}}{2+2\frac{A^2G}{(m+k_j)^2}}.
	\end{equation}
	$\lambda$ is error rate observed in parameter estimation, $A=\max\{m,k_j\}$ and $G=\frac{m+k_j}{mk_j}\ln\frac{m+k_j}{2\pi mk_j\lambda(1-\lambda)\epsilon^2}$.
	In summary, the extractable key length given by Eq.~(\ref{QSSlength}) guarantees the $\epsilon_s$-secrecy of our QSS protocol, which completes the proof. 
	\end{proof}

	\begin{table}[tp!]
		\caption{Different clicks to identify $\ket{\Phi^{+}_0}$ and $\ket{\Phi^{-}_0}$. In this table, we show the corresponding clicks on $V$ to identify $\ket{\Phi^{+}_0}$ and $\ket{\Phi^{-}_0}$ when $n$ is odd and even.}
		\begin{tabular}{ccc}
			\hline
			\hline
			&$n$ is odd&$n$ is even\\
			\hline
			$\ket{\Phi^{+}_0}$&even number of clicks&odd number of clicks\\
			$\ket{\Phi^{-}_0}$&odd number of clicks&even number of clicks\\
			\hline
			\hline
		\end{tabular}	\label{tab1}
	\end{table}
	
	\section{GHZ analyzer based on linear optical elements} \label{GHZanalyzer}

	\begin{figure}[tp!]
		\includegraphics[width=7.5cm]{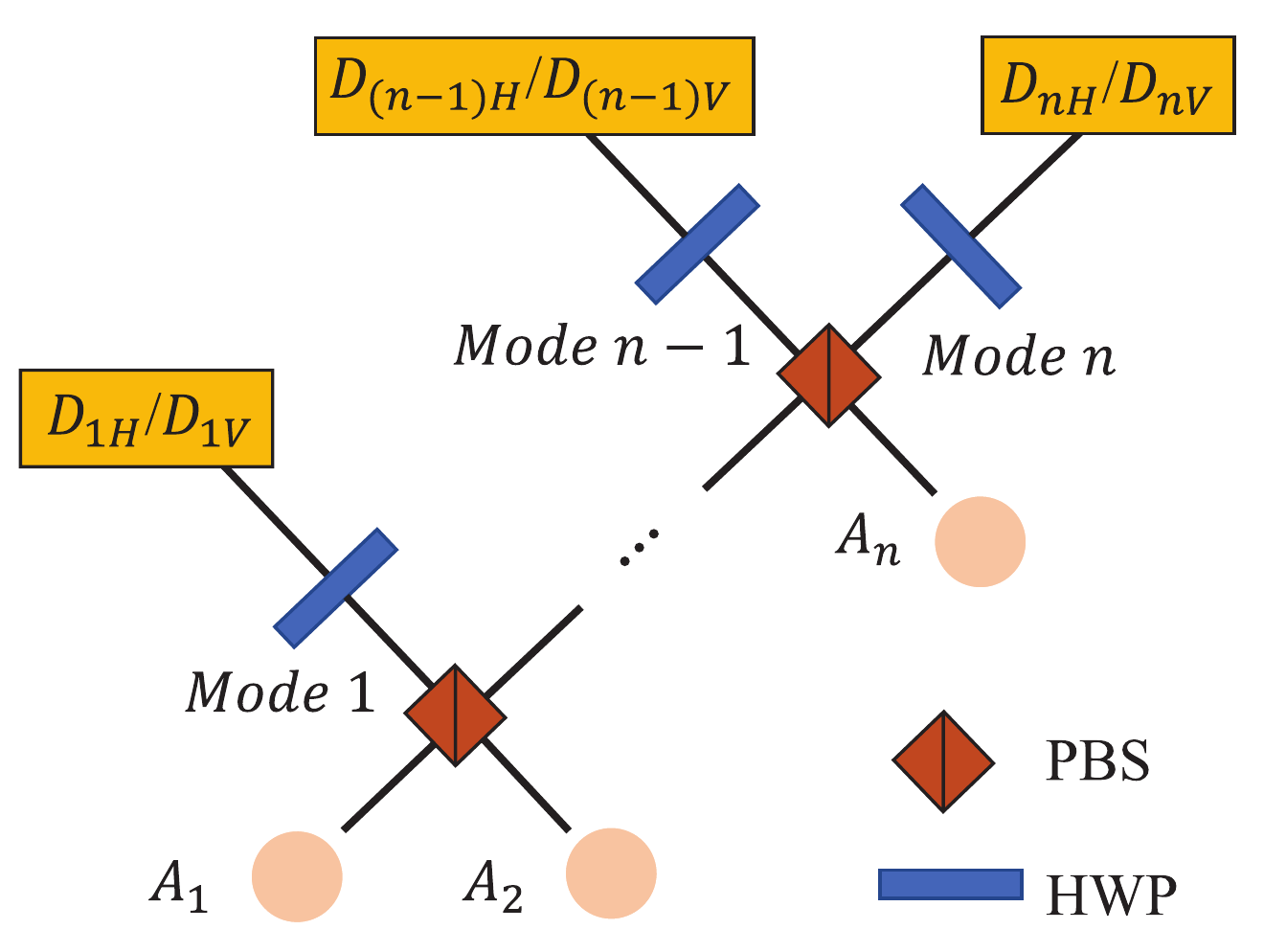}
		\caption{Schematic of Greenberger-Horne-Zeilinger (GHZ) analyzer based on linear optical elements. $\{A_i\}_{i=1,...,n}$: input modes; PBS: polarizing beam splitter which transmits $\ket{H}$ and reflects $\ket{V}$ polarizations; HWP: half-wave plate used to conduct a $45^{\circ}$ rotation of polarization. $D_{iH}/D_{iV}$ means detecting the $i$th mode in the Z basis.  
		}\label{fig3}
	\end{figure}
	
	The GHZ analyzer based on linear optical elements~\cite{pan1998greenberger}, as shown in Fig.~\ref{fig3}, is composed of just polarizing beam splitters (PBSs) and half-wave plates (HWPs) and can identify two of the $n$-particle GHZ states.
	We now explain how $n$-particle GHZ state $\ket{\Phi^{\pm}_0}=1/\sqrt{2}(\ket{HHH...H}\pm\ket{VVV...V})$ evolves in such analyzer.
	
	Suppose that $n$ particles of $\ket{\Phi^{\pm}_0}$ enter the GHZ analyzer shown in Fig.~\ref{fig3} each one through mode $A_i$ respectively and we express the input state using creation operator as 
	\begin{equation}
		\begin{split}
			\ket{\Phi^{\pm}_0}=&\frac{1}{\sqrt{2}}\left( \ket{H...H}_{A_1...A_n}\pm\ket{V...V}_{A_1...A_n}\right) \\
			=&\frac{1}{\sqrt{2}}\left( a^{H\dagger}_{A_1}...a^{H\dagger}_{A_n}\pm a^{V\dagger}_{A_1}...a^{V\dagger}_{A_n}\right) \ket{0}_{A_1...A_n}.
		\end{split}\label{inputstate}
	\end{equation}
	Here $a^{X\dagger}_{A_i}(X=H,V;i=1,...,n)$ represents the creation operator with $X$ polarization from mode $A_i$ and $\ket{0}$ is vacuum state.
	The polarizing beam splitter transmits $\ket{H}$ and reflects $\ket{V}$ polarization, where a phase of $\frac{\pi}{2}$ will be added on the output state.
	Therefore, we can find how $\ket{\Phi^{\pm}_0}$ evolves right after $n$ photons pass through PBS and before they enter HWP:
	\begin{equation}\label{afterPBS}
		\begin{split}
			\ket{\Phi^{\pm}_0}&\overset{\rm PBS}{\longrightarrow}\frac{1}{\sqrt{2}}\left( \aH{1}...\aH{n}\pm i^{2n-2}\aV{1}...\aV{n}\right) \ket{0}_{1...n}\\
			&=\frac{1}{\sqrt{2}}\left( \aH{1}...\aH{n}\pm (-1)^{n-1}\aV{1}...\aV{n}\right) \ket{0}_{1...n},
		\end{split}
	\end{equation}
	where $i$ is the imaginary unit and $a^{X\dagger}_{k}(X=H,V;k=1,...,n)$ represents the creation operator with $X$ polarization in $Mode$ $k$ shown in Fig.~\ref{fig3}.
	From Eq.~(\ref{afterPBS}), one can observe $n$-fold coincidences, which distinguishes $\ket{\Phi^{\pm}_0}$ from other $n$-particle GHZ states.
	
	Furthermore, after passing through the HWP, we can obtain 
	\begin{widetext}
	\begin{equation}\label{afterHWP}
			\frac{1}{2^{(n+1)/2}}\left[ (\aH{1}+\aV{1})(\aH{2}+\aV{2})...(\aH{n}+\aV{n})\pm (-1)^{n-1}(\aH{1}-\aV{1})(\aH{2}-\aV{2})...(\aH{n}-\aV{n})\right] \ket{0}_{12...n},
	\end{equation}
	from which we can identify $\ket{\Phi^{+}_0}$ and $\ket{\Phi^{-}_0}$.
	Because of the existence of factor $(-1)^{n-1}$ in Eq.~(\ref{afterHWP}), in the following we will discuss different criteria to identify $\ket{\Phi^{+}_0}$ and $\ket{\Phi^{-}_0}$ when $n$ is odd or even.
	
	To be specific, when $n$ is odd, $\ket{\Phi^{+}_0}$ evolves into the following state 
	\begin{equation}\label{phi0+evolve}
			\frac{1}{2^{(n+1)/2}}\left[ (\aH{1}+\aV{1})(\aH{2}+\aV{2})...(\aH{n}+\aV{n})+
			(\aH{1}-\aV{1})(\aH{2}-\aV{2})...(\aH{n}-\aV{n})\right] \ket{0}_{12...n},
	\end{equation}
	while $\ket{\Phi^{-}_0}$ evolves into
	\begin{equation}\label{phi0-evolve}
			\frac{1}{2^{(n+1)/2}}\left[ (\aH{1}+\aV{1})(\aH{2}+\aV{2})...(\aH{n}+\aV{n})-\\
			(\aH{1}-\aV{1})(\aH{2}-\aV{2})...(\aH{n}-\aV{n})\right] \ket{0}_{12...n}.
	\end{equation}
	\end{widetext}

	From Eq.~(\ref{phi0+evolve}) (Eq.~(\ref{phi0-evolve})), we can conclude that only products of creation operators with even (odd) number of $V$ polarization remains, which corresponds to even (odd) number of $\{D_{iV}\}_{i=1,...,n}$ being clicked.
	
	When $n$ is even, it is evident that different clicks corresponding to $\ket{\Phi^{+}_0}$ and $\ket{\Phi^{-}_0}$ exchange compared to clicks when $n$ is odd.
	
	For easier reference, we summarize the aforementioned results in Table~\ref{tab1}.

	\section{The gain under asymptotic limit}\label{derivation}
	According to Sec.~\ref{performance}, we state that under asymptotic limit the gain can be written as 
	\begin{equation}\label{Q}
		Q_{X}=Q^{\text{GHZ}}_{X}\cdot p_{\text{QND}}\cdot \sqrt{\eta_{\text{channel}}}\cdot \eta_{\text{sps}}\cdot \eta_a.
	\end{equation}
	In this Appendix, we present a derivation of Eq.~\ref{Q}.
	
	Before the derivation, for simplicity, we denote $p_{\text{QND}}\cdot \sqrt{\eta_{\text{channel}}}\cdot \eta_{\text{sps}}\cdot \eta_a$ as $\eta$, which represents the success probability of photon arrive at the GHZ analyzer.
	We first recall the definition of the gain $Q_{X}=\bar{N}/M$ and consider the calculation of $\bar{N}$.
	From the definition of $\bar{N}$, we have 
	\begin{equation}
		\bar{N}=\sum_{n=0}^MnP_{n|M},
	\end{equation}
	where $P_{n|M}$ is the probability when $n$ groups are successfully projected on GHZ states with $M$ multiplexing and can be expressed as 
	\begin{equation}
		P_{n|M}=\sum_{l=n}^MB_{n|l}\left( Q^{\text{GHZ}}_{X}\right) p_{l|M}.
	\end{equation}
	$B_{n|l}(p)=C_M^kp^k(1-p)^{M-k}$ with $p=Q^{\text{GHZ}}_{X}$ is a binomial distribution representing the probability of $n$ successful GHZ measurements conditioned on the existence of $l$ groups.
	Here $C_M^k=\begin{pmatrix}
		M\\k
	\end{pmatrix}$.
	$p_{l|M} = 3B_{l|M}(\eta)\left[ \sum_{k=l}^MB_{k|M}(\eta)\right] ^2-2\left[ B_{l|M}(\eta)\right] ^3$ is the probability of not less than $l$ single photons from all three parties with $M$ multiplexing.
	By utilizing $lB_{l|M}(p)=MpB_{l-1|M-1}(p)$ for $l>0$ and $B_{k|M}(p)=(1-p)B_{k|M-1}(p)+pB_{k-1|M-1}$ for $0<k<M$~\cite{azuma2015allqkd}, considering the asymptotic behavior of the maximum of binomial distribution, we have 
	\begin{equation}
		\lim_{M\rightarrow\infty}\bar{N}= Q^{\text{GHZ}}_{X}\sum_{l=0}^Mlp_{l|M}=MQ^{GHZ}_{X}\cdot\eta.
	\end{equation}
	Therefore, we have 
	\begin{equation}
		\begin{split}
			Q_{X}=&\lim_{M\rightarrow\infty}\frac{\bar{N}}{M}=Q^{\text{GHZ}}_{X}\cdot\eta\\
			=&Q^{\text{GHZ}}_{X}\cdot p_{\text{QND}}\cdot \sqrt{\eta_{\text{channel}}}\cdot \eta_{\text{sps}}\cdot \eta_a.
		\end{split}
	\end{equation}
	
	\section{Estimation of the success probability of GHZ measurement and bit (phase) error rate}\label{estimation}
	
	\begin{figure}[tbp!]
		\includegraphics[width=8.5cm]{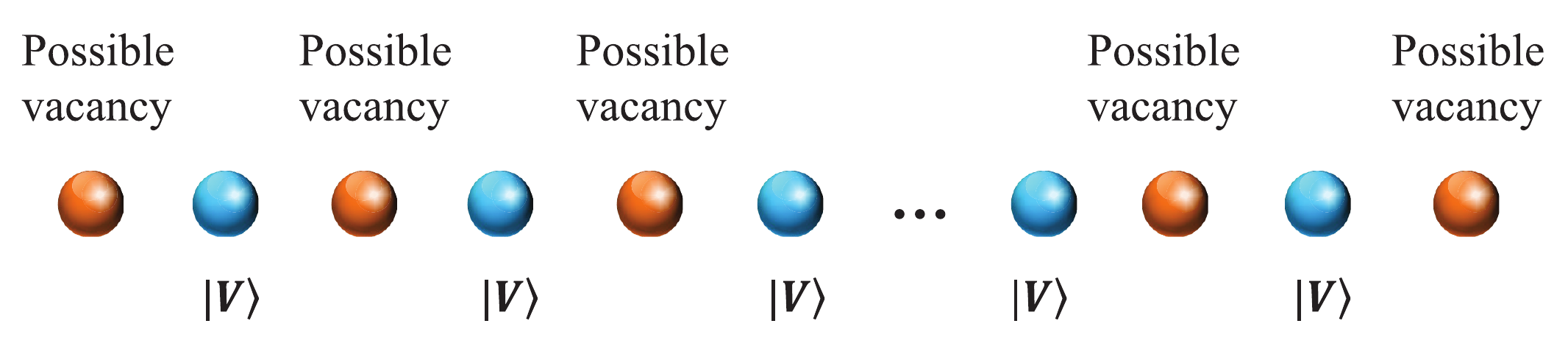}
		\caption{The arrangement of $\ket{V}$ photons and possible vacancies. We use blue and orange circles to denote $\ket{V}$ photons and vacancies. When inserting $\ket{H}$ photons into fixed $\ket{V}$ photons, we first determine the number of vacancies and then determine the number of $\ket{H}$ in each vacancy. Finally we obtain a distribution of input photon state.
		}\label{afig1}
	\end{figure}
	In this Appendix, we give the calculation of the gain and bit (phase) error rate of our protocol.
	We start with recalling the classical part of the preceding MDI-QSS~\cite{fu2015long}.
	After step ($iv$) of our protocol, $\{A_i\}_{i=1,..,n+1}$ postselect the events where they prepare the states with the same basis through an authenticated public channel.
	One should note that when all $n+1$ users choose the $X$ basis and the state is projected onto $\ket{\Phi^-_0}$, $A_1$ will perform a bit flip on his classical bit.
	Finally, all users estimate parameters through experiment and extract keys after classical error correction and privacy amplification.
	In the following we provide an explicit description of the calculation of $Q^{\text{GHZ}}_{X(Z)}$ and $E_{Z(X)}$.
	
	We first consider the calculation of $Q^{\text{GHZ}}_{Z}$ and $E_{Z}$.
	For simplicity, we introduce some notations as follows.
	$x_{0}$ refers to the probability of $D_{iH/V}$ clicking when vacuum state is in the $i$ th mode.
	$x_{1C (E)}$ refers to the probability of correct (erroneous) click when the single photon state is in the $i$th mode.
	Here the correct click means $D_{iH(V)}$ clicks when $\ket{H}(\ket{V})$ is input state and the meaning of erroneous click is $D_{iH(V)}$ clicks when $\ket{V}(\ket{H})$ inputs.
	$x_{2C (E)}$ refers to the probability of correct (erroneous) click when two photons are in the $i$ th mode.
	It is easy to calculate the probability of successful GHZ projection when $n$ users prepare state with perfect bit correlation in the $Z$ basis, $i$.$e$. $\ket{HH...H}$ and $\ket{VV...V}$.
	By considering the evolution of $\ket{HH...H}$ and $\ket{VV...V}$ in the GHZ analyzer shown in Fig.~\ref{fig3}, we have 
	\begin{equation}
		Q_{nH}=Q_{nV}=(x_{1C}+x_{1E})^n,
	\end{equation}
	where $Q_{nH}$ ($Q_{nV}$) is the success probability of GHZ projection when $\ket{HH...H}$ ($\ket{VV...V}$) inputs.
	
	Now we consider how to estimate $Q_{(n-k)H,kV}$ ($k\ge 1$), the sum of the gain when the input state owns $k$ photons in $V$ polarization.
	We limit $k\le n/2$ since for $k> n/2$ we have $Q_{(n-k)H,kV}=Q_{kH,(n-k)V}$ due to the symmetry.
	The calculation of $Q_{(n-k)H,kV}$ can be solved as a counting problem since the gain is different under various input arrangements in the following way.
	At first, we need to determine the distribution of $\ket{V}$ photons in $n$ modes.
	We assume that $\ket{V}$ photons are fixed and the other $\ket{H}$ photons are inserted into them which is shown in Fig.~\ref{afig1}.
	Such insertion can be finished in two steps.
	First, determine the number of vacancies where the $\ket{H}$ photons will be inserted into.
	Then we decide the number of $\ket{H}$ photons in each vacancy and we can get a distribution of input photons.
	One should note that for the GHZ analyzer used in this paper, choosing the leftmost vacancy in Fig.~\ref{afig1} is the same as choosing the rightmost vacancy.
	As a result, when two vacancies are chosen at the same time, they should be viewed as a single one vacancy. 
	We denote the number of all possible distributions when there is $l$ vacancies in $k$ photons in $\ket{V}$ as $g_k(l)$ and the corresponding success probability of GHZ projection as $f(l)$.
	Then we have the following expression:
	\begin{equation}
		Q_{(n-k)H,kV}=\frac{1}{2^n}\sum_{l=1}^kg_k(l)f(l),
	\end{equation} 
	where $g_k(1)=n$ and for $l\neq1$
	\begin{equation}
		g_k(l)=
			\left[\frac{1}{(l-1)!}(C_{k+1}^{l}-C_{k-1}^{l-2})+\frac{1}{l!}C_{k-1}^{l-1}\right]\frac{(n-k-1)!}{(n-k-l)!}.
	\end{equation}
	In addition, we have
	\begin{equation}
		f(l)=2^l(x_{2C}+x_{2E})^lx_0^l(x_{1C}+x_{1E})^{n-2l}.
	\end{equation}
	Here and in the following we define $C_m^n=\begin{pmatrix}
		m\\n
	\end{pmatrix}$.
	We now make a remark on critical situation.
	When $n$ is even and $l=k=n/2$, we have 
	\begin{equation}
		g_{\frac{n}{2}}(\frac{n}{2})=\frac{2(n-k-1)!}{(l-1)!(n-2k)!}.
	\end{equation}
	In summary, we present the following expression:
	\begin{equation}
		Q^{\text{GHZ}}_{Z}=2Q_{nH}+\sum_{k=1}^{n-1}Q_{(n-k)H,kV}.
	\end{equation}
	According to the definition of errors under $Z$ basis, we have 
	\begin{equation}
		E_Z=\frac{1}{Q^{\text{GHZ}}_{Z}}\sum_{k=1}^{n-1}Q_{(n-k)H,kV}.
	\end{equation}

	Now we consider the gain $Q_X^{\text{GHZ}}$ and phase error rate $E^X$.
	Due to the equality of density matrix, we can directly conclude that $Q_X^{\text{GHZ}}=Q_Z^{\text{GHZ}}$.
	To estimate the phase error rate, we need to calculate the success probability of projection on $\ket{\Phi_0^+}$ and $\ket{\Phi_0^-}$ respectively.
	We decompose states prepared in $X$ basis into $Z$ basis and aforementioned methods can be used.
	We summarize the following results according to the evaluation of the states in the GHZ analyzer.
	
	First we consider the situation when there is even number of $\ket{-}$ photons.
	If $n$ is odd, $Q_{\text{even}}^{\Phi_0^+}$ ($Q_{\text{even}}^{\Phi_0^-}$), the success probability of projection on $\ket{\Phi_0^+}$ ($\ket{\Phi_0^-}$) can be given by
	\begin{widetext}
	\begin{equation}
		\begin{split}
			Q_{\text{even}}^{\Phi_0^+}=&\frac{1}{4^{n-1}}\sum_{i=0}^{(n-1)/2}\sum_{k=0}^{(n-1)/2}C_n^{2i}C_n^{2k}x_{1E}^{2k}x_{1C}^{n-2k}+
			\frac{1}{2^n}\sum_{k=1}^{(n-1)/2}\sum_{l=1}^{k}g_k(l)f(l),\\
		Q_{\text{even}}^{\Phi_0^-}=&\frac{1}{4^{n-1}}\sum_{i=0}^{(n-1)/2}\sum_{k=0}^{(n-1)/2}C_n^{2i}C_n^{2k+1}x_{1E}^{2k+1}x_{1C}^{n-2k-1}+\frac{1}{2^n}\sum_{k=1}^{(n-1)/2}\sum_{l=1}^{k}g_k(l)f(l).
		\end{split}
	\end{equation}	
	If $n$ is even, we have
	\begin{equation}
		\begin{split}
			Q_{\text{even}}^{\Phi_0^+}=&\frac{1}{4^{n-1}}\sum_{i=0}^{n/2-1}\sum_{k=0}^{n/2}C_n^{2i+1}C_n^{2k}x_{1E}^{2k}x_{1C}^{n-2k}+\frac{1}{2^n}\sum_{k=1}^{n/2}\sum_{l=1}^{k}g_k(l)f(l);\\
			Q_{\text{even}}^{\Phi_0^-}=&\frac{1}{4^{n-1}}\sum_{i=0}^{n/2-1}\sum_{k=0}^{n/2-1}C_n^{2i+1}C_n^{2k+1}x_{1E}^{2k+1}x_{1C}^{n-2k-1}+\frac{1}{2^n}\sum_{k=1}^{n/2}\sum_{l=1}^{k}g_k(l)f(l).
		\end{split}
	\end{equation}

	Then we consider the situation when there is odd number of $\ket{-}$ photons.
	If $n$ is odd, $Q_{\text{odd}}^{\Phi_0^+}$ and $Q_{\text{odd}}^{\Phi_0^-}$ can be given by
	\begin{equation}
		\begin{split}
			Q_{\text{odd}}^{\Phi_0^+}=&\frac{1}{4^{n-1}}\sum_{i=0}^{(n-1)/2}\sum_{k=0}^{(n-1)/2}C_n^{2i+1}C_n^{2k}x_{1E}^{2k+1}x_{1C}^{n-2k-1}+\frac{1}{2^n}\sum_{k=1}^{(n-1)/2}\sum_{l=1}^{k}g_k(l)f(l);\\
			Q_{\text{odd}}^{\Phi_0^-}=&\frac{1}{4^{n-1}}\sum_{i=0}^{(n-1)/2}\sum_{k=0}^{(n-1)/2}C_n^{2i+1}C_n^{2k+1}x_{1E}^{2k}x_{1C}^{n-2k}+\frac{1}{2^n}\sum_{k=1}^{(n-1)/2}\sum_{l=1}^{k}g_k(l)f(l),
		\end{split}
	\end{equation}
	If $n$ is an even number, we have
	\begin{equation}
		\begin{split}
			Q_{\text{odd}}^{\Phi_0^+}=&\frac{1}{4^{n-1}}\sum_{i=0}^{n/2}\sum_{k=0}^{n/2-1}C_n^{2i}C_n^{2k+1}x_{1E}^{2k+1}x_{1C}^{n-2k-1}+\frac{1}{2^n}\sum_{k=1}^{n/2}\sum_{l=1}^{k}g_k(l)f(l);\\
			Q_{\text{odd}}^{\Phi_0^-}=&\frac{1}{4^{n-1}}\sum_{i=0}^{n/2}\sum_{k=0}^{n/2}C_n^{2i}C_n^{2k}x_{1E}^{2k}x_{1C}^{n-2k}+\frac{1}{2^n}\sum_{k=1}^{n/2}\sum_{l=1}^{k}g_k(l)f(l),
		\end{split}
	\end{equation}
	
	Based on the results above and the definition of error under $X$ basis, we can express the phase error rate as 
	\begin{equation}
		E_X=
		\left\{
		\begin{split}
			\frac{1}{2^nQ_X^{\text{GHZ}}}&\sum_{m=0}^{n/2}C_{n}^{2k}Q_{\text{even}}^{\Phi_0^-}+\sum_{m=0}^{n/2-1}C_{n}^{2k+1}Q_{\text{odd}}^{\Phi_0^+},n \text{ is even},\\
			\frac{1}{2^nQ_X^{\text{GHZ}}}&\sum_{m=0}^{(n-1)/2}C_{n}^{2k}Q_{\text{even}}^{\Phi_0^-}+\sum_{m=0}^{(n-1)/2}C_{n}^{2k+1}Q_{\text{odd}}^{\Phi_0^+},n \text{ is odd}.
		\end{split}
		\right.
	\end{equation}

	Finally using the above equations we can estimate the key rate of our QSS.
	\end{widetext}	


%

\end{document}